\documentclass[prb,twocolumn,groupedaddress,showpacs]{revtex4}
\usepackage{epsfig}
\usepackage{amsmath}
\usepackage{amsfonts}

\newcommand{\be}{\begin{equation}}
\newcommand{\ee}{\end{equation}}
\newcommand{\bea}{\begin{eqnarray}}
\newcommand{\eea}{\end{eqnarray}}
\def\nn{\nonumber\\}
\newcommand{\la}{\langle}
\newcommand{\ra}{\rangle}
\newcommand{\ham}{{\cal H}}
\newcommand{\p}{\partial}
\newcommand{\s}{\sigma}

\begin{document}

\title{Spinless Fermionic ladders in a magnetic field: Phase Diagram}

\author{Sam T. Carr}
\affiliation{The Abdus Salam ICTP, Strada Costiera 11, 
Trieste, 34100, Italy.}
\author{Boris N. Narozhny}
\affiliation{The Abdus Salam ICTP, Strada Costiera 11, 
Trieste, 34100, Italy.}
\author{Alexander A. Nersesyan}
\affiliation{The Abdus Salam ICTP, Strada Costiera 11, 
Trieste, 34100, Italy.}
\affiliation{The Andronikashvili Institute of Physics, Tamarashvili 6,
0177, Tbilisi, Georgia}

\date{\today}

\begin{abstract}

 The system of interacting spinless fermions hopping on a two-leg
 ladder exhibits a series of quantum phase transitions when subjected to
 an external magnetic field. At half filling, these are either $U(1)$
 Gaussian phase transitions between two phases with distinct types of
 long-range order or Berezinskii-Kosterlitz-Thouless transitions
 between ordered and gapless phases.

\end{abstract}


\maketitle

\section{Introduction}

The behavior of interacting electron systems under the action of an
external magnetic field, affecting orbital motion of the particles, is
a subject of intense research of the last several decades. It has long
been established that the magnetic field has a dramatic effect on all
properties of the system. Even in the absence of interaction, the
spectrum of the free Fermi gas is modified and exhibits Landau
quantization\cite{lan} in the continuum or the Hofstadter
spectrum\cite{hof} on a two-dimensional (2D) lattice. The most
spectacular consequence of this phenomenon is the Quantum Hall Effect
(QHE).\cite{qhe} Electron-electron interaction compounds the
complexity of the problem, giving rise to the Fractional Quantum Hall
Effect (FQHE).\cite{fhe}

The current theoretical understanding of the effect of the magnetic
field on the properties of electron systems was achieved by a
combination of various methods and techniques, each of which is
strictly speaking only applicable in a certain parameter range. What
is still lacking is a comrehensive approach that would unify all of
the different aspects of the problem into a single coherent picture.
Perhaps at present such a goal is too ambitious. However, as a first
small step in this direction, one can ask whether such a comprehensive
approach can be formulated for a simpler model, that would on one hand
be a problem of interacting electrons in the magnetic field and on the
other hand would, at least in principle, allow a generalization
towards the original problem.

The simplest model of interacting fermions that incorporates orbital
effects of an external magnetic field is that of spinless fermions
hopping on a two-leg ladder. This model is simple enough not to
exhibit the multitude of small gaps in the single-particle spectrum
(characteristic of the Hofstadter problem\cite{hof}). Yet the magnetic
flux piercing the plaquettes of the ladder changes the ground state
properties of the system giving rise to non-trivial phases and
inducing quantum phase transitions.

Ladder models\cite{thl,fab} occupy a special place in the field of
strongly correlated electron systems. On one hand, they describe (at
least within some range of temperatures) behavior of many naturally
found compounds\cite{ex1}, including carbon nanotubes,\cite{ex2,th4}
as well as artificially manufactured structures.\cite{ex3} On the
other hand, they provide a fertile ground for application of
theoretical techniques developed for one-dimensional systems\cite{th1}
- i.e. non-perturbative approaches leading to asymptotically exact
results. Furthermore, they are a first step for various attempts at
generalization of the lore of physics in one spatial dimension to
higher-dimensional problems.\cite{th2}

As is often the case, situations where the number of particles is
commensurate with the lattice attract the most attention since only
then can long range order (LRO) develop. We have previously
shown\cite{us} that at a quarter filling, the presence of the flux
results in exciting effects which do not exist in the absence of the flux. In
particular the uniform external magnetic field can lead to a staggered
flux (or orbital antiferromagnet) phase, which furthermore has
fractionally charged excitations.

The purpose of the present paper is to describe the full phase diagram
of interacting spinless fermions on the two-leg ladder at
$1/2$-filling in the presence of an external magnetic field. To drive
a system to criticality by applying the magnetic field is an
intriguing possibility which should be more accessible in experiment
than varying coupling constants. The phase diagram displays a
multitude of quantum phase transitions induced by the flux. There are
two types of these phase transitions: (i)
Berezinskii-Kosterlitz-Thouless (BKT) transitions\cite{bkt} between
ordered and disordered ground states, and (ii) $U(1)$ Gaussian phase
transitions between different ordered ground states. Here we choose
the fermions to be spinless in order to eliminate Zeeman splitting and
focus on the orbital effects of the magnetic field. The case of
spin-1/2 particles will be discussed in a separate publication.

Traditionally \cite{gia,boz} ladder models have been treated in two
complimentary approaches. On one hand, one can start with the model of
two decoupled chains, define the low energy effective theory for each
of them, and then treat both the single-particle transverse hopping
and two-particle inter-chain correlations
perturbatively.\cite{nlk93} On the other hand, one could start with
the exact single particle basis (given by two bands) and then proceed
with the corresponding low energy limit.\cite{fab} For the major part
of this paper we will be using the latter approach which allows us to
treat intra- and inter-chain processes on equal footing. However, it
is well known \cite{and} (at least in the absence of the magnetic
field) that for weak enough inter-chain tunneling there exists the
phenomenon of Anderson confinement, i.e. the suppression of the
inter-chain single-particle tunneling by intra-chain two-particle
correlations. This effect can be seen in either picture, but it is
more intuitive to discuss it within the chain approach. Both ways
should produce the same physical results, but understanding the
relation between the two approaches allows us to establish the limits
of applicability of the effective low energy theories that can be
derived within either picture. Moreover, we discuss how the Anderson
confinement regime is affected by the magnetic flux.

The remainder of the paper is organized as follows. We start by
defining the microscopic Hamiltonian of the model and proceed directly
to the results, discussing the phase diagram and the other physical
properties of the model. Then we outline the details of the
calculations within the weak coupling (bosonization) approach. In
Section~\ref{band} we derive the effective low energy theory in the
band picture. In Section~\ref{chain} we turn to the chain picture and
discuss its relation to the band approach. Section~\ref{strong} is
devoted to the strong coupling limit of our model and is followed by a
brief summary of the results. Mathematical details are relegated to
Appendices.

\section{Model and results}
\label{intro}

In this section we present our results. We start by defining the
microscopic Hamiltonian of our model and proceed to discuss the
zero-temperature phase diagram.

\subsection{The Hamiltonian}

We consider a tight-binding model of spinless fermions on a two-leg
ladder described by the Hamiltonian 
\bea 
H = &-&
\frac{1}{2}\sum_{i\s}\left[ t_{\parallel}(\s) c_{i,\sigma}^\dagger
c_{i+1,\sigma} + h.c. \right] - t_{\perp} \sum_{i\sigma}
c_{i,\sigma}^\dagger c_{i,-\sigma} \nn &+& U \sum_i n_{i+} n_{i-} +
V\sum_{i\sigma} n_{i,\sigma} n_{i+1,\sigma}.
\label{ham}
\eea
\noindent
Here $c_{i,\sigma}$ is the electron annihilation operator on the chain
$\sigma=\pm$ at the site $i$; $n_{i\s} =
c^{\dagger}_{i,\sigma}c_{i,\sigma}$ are the occupation number
operators; $t_\perp$ and $t_{\parallel}$ are the transverse and
longitudinal hopping amplitudes, respectively. The last two terms in
Eq.~(\ref{ham}) describe nearest neighbour inter- and intra-chain
interactions. The chosen form of short-range interaction is quite
representative because it reflects the generic symmetry of the ladder
and yields the most general effective field theory in the low-energy
limit. Our notation reflects the well-known analogy between ladder
models of spinless fermions and Hubbard-like chains of spin-1/2
particles. In our case however, the $SU(2)$ symmetry is explicitly
broken in the Hamiltonian~(\ref{ham}) by the inter-chain hopping and
the $V$ interaction term (the former is analogous to a Zeeman energy
due to an external magnetic field alng the $x$-axis, while the latter
is a counterpart of an exchange anisotropy along the $z$-axis).

The external magnetic field $B$ is introduced by means of the Peierls
substitution.\cite{pei} In the Landau gauge \cite{dau} with the
vector potential ${\bf A} = B(-y, 0, 0)$ the transverse hopping term
is independent of the field, while the longitudinal hopping amplitude
can be written as
\be
t_{\parallel}(\sigma) = t_0 e^{i \pi \sigma  f},
\label{t}
\ee
\noindent
where $f$ is the magnetic flux through the elementary plaquette in
units of flux quantum $\phi_0 = hc/e$. Expressed in terms of the flux
the model is explicitly gauge invariant.

\begin{figure}[ht]
\vspace{0.1 cm}
\epsfxsize=8.1 cm
\centerline{\epsfbox{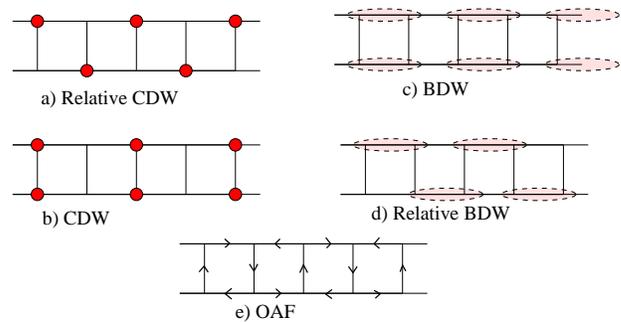}}
\caption{Cartoon depictions of possible ordered phases at $1/2$
filling. The dots represent excess fermion dimerization on the
sites. The ellipses represent excess occupation on the bonds. The
arrows represent local currents. Note that the OAF and the BDW
coexist - see Section~\ref{band} for an explanation.}
\label{halfphases}
\end{figure}

\subsection{Phase Diagram}
\label{phased}

At half filling (one fermion per rung) the model (\ref{ham}) is
characterized by a rather rich phase diagram. Depending on the values
of microscopic parameters the ground state of the model may posess
true long range order. Possible ordered phases are illustrated in
Fig.~\ref{halfphases}. The cartoons show a strong-coupling picture of
the phases: charge density waves (CDW), where particles are localized
on sites of the ladder; bond density waves (BDW) with dimerized links
along the chains; and the orbital antiferromagnet\cite{nlk93} (OAF),
sometimes referred to as the staggered flux phase or a d-density wave,
where the particle density remains uniform, but there exist
non-vanishing local currents that have opposite directions on
alternate bonds. Notice that, in the spin language, the Relative CDW is
similar to a spin density wave (SDW) with spins polarized along $z$ (a
Neel state SDW$^z$), whereas the OAF is equivalent to a SDW$^y$.

In addition, the model (\ref{ham}) allows for various phases that do
not posess long-range order. Using the aforementioned analogy with the
Hubbard chain, we may discuss the model (\ref{ham}) in terms of
``spin'' (or ``relative'') and ``charge'' (or ``total'') sectors.  In
Section~\ref{int1} we show that, in the low-energy limit, the
``charge'' and ``relative'' sectors of the model asymptotically
decouple. In all of the ordered phases both sectors are
gapped. However, it is possible to have a phase where only one of the
sectors acquires a spectral gap. Phases where only the ``charge''
sector is gapped, irrespective of the type of dominant correlations,
we will call the Mott Insulator (MI). The cases where the gap exists
in the ``relative'' sector only will be called the Luther-Emery Liquid
(LEL).\cite{lel} Finally, when both sectors are gapless, the system
represents a Luttinger Liquid (LL).

{
\begin{figure}[ht]
\epsfxsize=7 cm
\centerline{\epsfbox{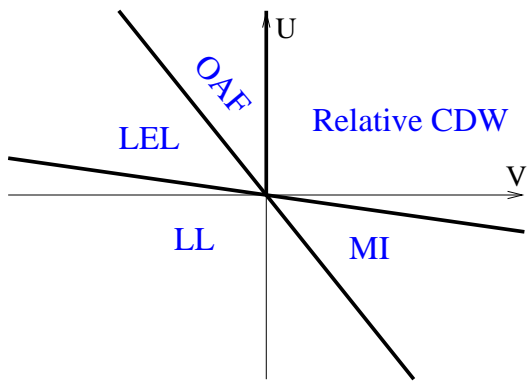}}
\caption{Phase diagram at $B=0$ (after Ref.~\onlinecite{boz}). Phase
boundaries correspond to the lines $V=0$, $U/2V = - 2 + \tau^2$ and
$U/2V = - \tau^2$.}
\label{noflux}
\end{figure}
}

The phase diagram in the absense of the flux is known.\cite{gia,boz}
For the sake of clarity we include it in Fig.~\ref{noflux}. This
phase diagram is valid for suffiently large values of $t_{\perp}$
where delocalization of the fermions across the rungs suppresses the
CDW phase (which happens in the absence of the inter-chain tunneling).
There are two ordered phases: (i) for purely repulsive interactions
$U,V>0$, one has a relative CDW as to be expected (see
Section~\ref{strong}); (ii) for repulsive interchain interaction and
not too strong attractive in-chain interaction, the ground state is
the orbital antiferromagnet.\cite{ner,nlk93} The phase diagram in
Fig.~\ref{noflux} was obtained within a weak-coupling bosonization
approach. The phases do exist when the coupling becomes strong,
however the exact location of the phase boundaries might change.

Once the magnetic field is applied, the system may exhibit additional
phase transitions. In Fig.~\ref{wcpd} we plot the entire weak-coupling
phase diagram for the model at half-filling and sufficiently large
$t_{\perp}$ (see next subsection).  The magnetic flux
varies along the vertical axis, so that the diagram in
Fig.~\ref{noflux} corresponds to the bottom axis of
Fig.~\ref{wcpd}. The ratio of the microscopic interaction parameters
of the Hamiltonian~(\ref{ham}), $U/2V$, is represented in
Fig.~\ref{wcpd} through the angular variable $\theta = \tan^{-1} U/2V$
which varies along the horizontal axis. Four different regions
corresponding to various signs of the constants $U$ and $V$ are
indicated on the upper horizontal line in Fig.~\ref{wcpd}. The
analytic description of the phase boundaries, based on the weak
coupling theory, is given in Appendix~\ref{phasebd}. The position of
the boundaries depends on the applied field, the ratio $U/2V$
(i.e. $\theta$) and the ratio of the hopping parameters, $\tau =
t_\perp/t_0$. We plot the phase diagram as a function of $\theta$ and
flux for $\tau=0.25$, the value is arbitrary but representative as
long as $\tau$ exceeds a possible gap in the ``relative'' sector.  At
other values of $\tau$ the topology of the phase diagram and
classification of the phases do not change qualitatively. Similarly,
if we modify our model~(\ref{ham}) to include other short-range
interaction terms, the only effect on the phase diagram would again be
just the shift of the phase boundaries.

Let us now describe the phase transitions induced by the applied field
(i.e. the vertical direction in Fig.~\ref{wcpd}).  We assume that the
inter-chain hopping parameter $\tau$ is not too small (see next
subsection). Note, that since the model~(\ref{ham}) is invariant under
the transformation $f \rightarrow 1-f$, $\sigma \rightarrow
-\sigma$,we only need to consider the flux within the range $0<f<1/2$.
Moreover, when the flux is large enough, $\sin^2\pi f>1-\tau^2$, there
is a band gap in the single particle spectrum of the model. That state
is largely unaffected by interaction effects (at least within the
weak-coupling limit), and thus we will restrict our discussion to
smaller values of $f$. While the weak-coupling approach can not be
trusted at fields too close to the band gap limit (since the Fermi
velocity becomes too small), we continue the phase boundaries up to
that point. All phase transitions of interest happen sufficiently far
from that region. Apart from a brief discussion in Sec.~\ref{strong},
we will not consider the details of the transition to the band
insulator in this paper.

The most interesting features of the phase diagram in Fig.~\ref{wcpd}
are a sequence of $U(1)$ phase transitions between different ordered
states and reentrant transitions. Understanding of these transitions
is based on the fact that, as shown in Section~\ref{int1}, in the
low-energy limit the ``charge'' and ``relative'' degrees of freedom of
the model decouple, and each sector is described by a sine-Gordon
model (see Eq.~(\ref{halfbos}) and Appendix~\ref{phasebd}). Phases
with LRO correspond to strong-coupling regimes in both sectors. The
phases whose order parameters are mapped onto each other under a sign
change of the corresponding coupling constant (the amplitude of the
cosine term) are mutually dual. The associated U(1) Gaussian
criticality occurs at the self-dual lines, i.e. when the one of those
coupling constants vanish. Such a duality is commonplace in low-energy
effective theories, indeed more complicated non-Abelian dualities
were found recently in the SU(4) Hubbard model.\cite{lab04}  
However, it has also been recently shown\cite{momoi04}
that for certain ladder models the (Abelian) duality between different phases
turns out to be not only a symmetry emerging in the low-energy limit
but an exact property of the underlying microscopic model.

{
\begin{figure*}
\begin{center}
\epsfig{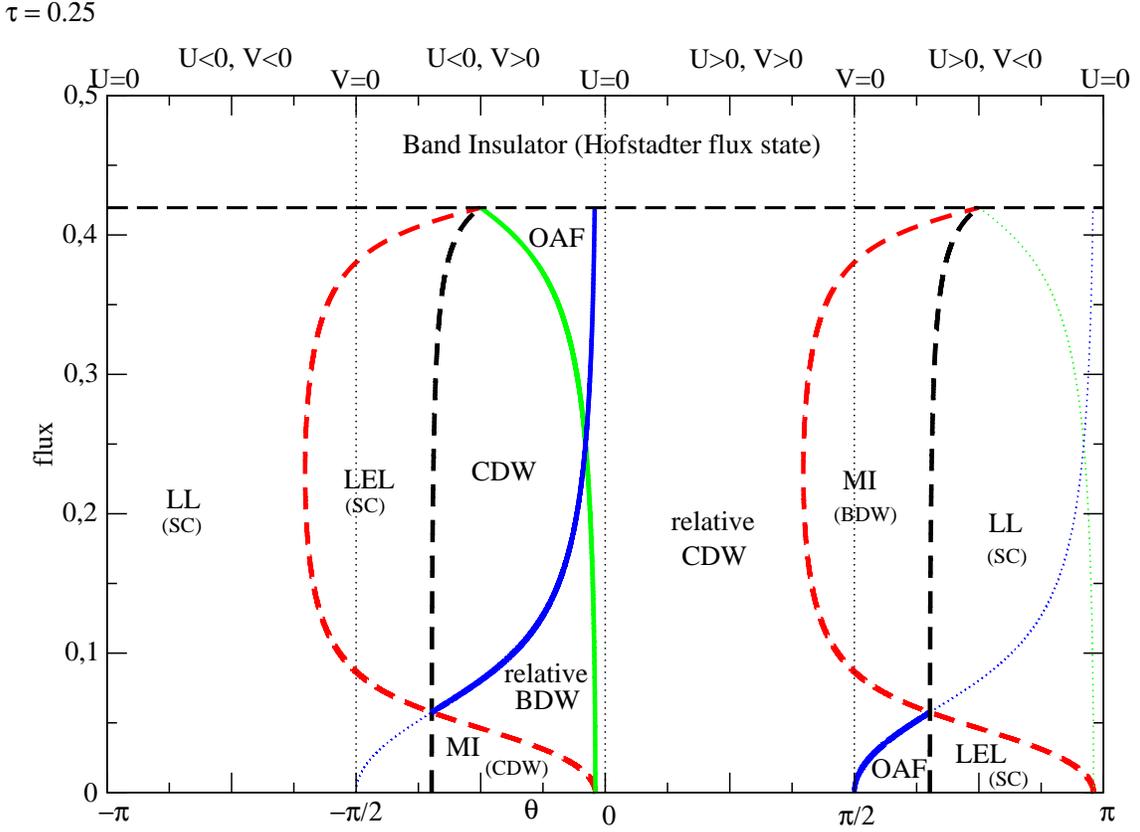}
\end{center}
\caption{The weak-coupling phase diagram in the magnetic field. We
plot the flux along the vertical axis and the angle $\theta$ (defined
as $\theta = \tan^{-1} U/2V$) along the horizontal axis. As the phases
depend on the signs of the interaction parameters, they are indicated
at the top of the diagram. Ordered phases are illustrated pictorially
in Fig.~\ref{halfphases}. The corresponding order parameters are
listed in Table~\ref{op}. The disordered phases are characterized by
dominant correlations (indicated in parentheses). For large values of
the flux ($\sin^2\pi f>1-\tau^2$), there is a band gap in the
non-interacting picture. The thick solid lines (blue and green online)
represent $U(1)$ Gaussian transitions between mutually dual ground
states with long-range order, and the thick dotted lines (black and
red online) are Berezinski-Kosterlitz-Thouless phase transitions
corresponding to opening of a gap in one of the sectors. }
\label{wcpd}
\end{figure*}
}

Gaussian transitions occur in two domains of the phase diagram: (i)
for repulsive intrachain interaction ($V>0$) and weak attractive
interchain interaction ($U<0$), and (ii) for $U>0$ and weak $V<0$. In
the first case (i), at zero flux the system is a MI. The dominant
(longest-range) correlation turns out to be that of the $2k_F$
component of the total charge density (hence the label ``(CDW)'' in
Fig.~\ref{wcpd}; see Section~\ref{nop} for details on dominant
correlations). As the flux is increased, the system becomes a relative
BDW (via the BKT transition where a gap opens in the ``relative''
sector of the effective theory). Further increase of $f$ drives the
system through the $U(1)$ transition to a CDW phase. At larger values
of the flux (approaching the band gap limit) values of flux, the
system eventually becomes an OAF, again through the Gaussian
transition.

The second transition (ii) occurs at small values of the flux. The
zero-field ground state is an OAF. As we turn on the flux, the system
undergoes a $U(1)$ transition towards a relative CDW state. Further
increase of the flux results in closing of the gap in the relative
sector (via the BKT transition) and the system becomes a MI, but now
the dominant correlation is that of the $2k_F$ component of the
transverse bond density, labeled by ``(BDW)''.

This latter transition turns out to be reentrant. As the flux is
further increased the system returns (again via the BKT transition)
back to the relative CDW state. There is another example of a
reentrant transition in the phase diagram, for $U<0$ and small $V$ the
zero-field ground state is a Luttinger Liquid, which once subjected to
the external field first becomes a LEL by opening a gap in the
``relative'' sector, and then at higher field comes back to the LL
state in which the most singular fluctuations are those of the pairing
operator at momentum $\pi-2k_F$. In the LEL phase the ``relative''
sector is gapped and the ``charge'' sector is characterized by the
dominant correlation of the pairing operator at zero momentum.

Finally there is a large part of the phase diagram which is robust
against the application of the external field. When both inter- and 
intra-chain interactions are attractive, the LL (that is the zero-field
ground state) is mostly unaffected by the field. More interesting is 
the situation when both interactions are repulsive. The zero-field 
ground state is the relative CDW. It turns out that this long-range order
survives under the application of the field (except possibly for the 
transition to the MI for weak $V$ discussed above).

\subsection{Commensurate-Incommensurate transition}
\label{cim}

The above phase diagram breaks down if the parameter $h = [\sin^2 (\pi
f) + \tau^2]^{1/2}$, which in the noninteracting case determines the
splitting between the Fermi momenta of different bands, is too small
(see Sections~\ref{band} and \ref{chain}). Then, the part of the phase
diagram that corresponds to attractive inter-chain interaction $U<0$
acquires additional ordered phases. This is the result of additional
inter-band scattering processes that at larger $h$ violate momentum
conservation in the low energy effective theory (based on the two-band
description). The latter issue reveals the dichotomy between the two
starting points already mentioned in the Introduction: chain basis
versus band basis.  If one starts with a solution (however complete)
for two independent chains and then tries to take into account the
inter-chain hopping (as well as the flux) in perturbation theory, then
the processes mentioned above are present in the theory from the
beginning.  In the case when these processes generate a gap in the
spectrum of relative degrees of freedom, a finite splitting of the
Fermi momenta would not take place unless the parameter $h$ exceeds
its critical value comparable with the gap. This is the well-known
commensurate-incommensurate transition.\cite{boz} As the parameter $h$
increases further, the ``two-chain'' approach fails because
renormalization of the parameters of the theory becomes sizable at
sufficiently large $h$. At that point one would be forced to start
with the exact, two-band single-particle spectrum of the ladder.
However, this would seemingly neglect the processes in question as
they appear to violate momentum conservation.

In Section~\ref{chain} we discuss the relation between the two
approaches to ladder problems and show that if one uses either
approach properly, then the final result is independent of the
starting point, as should be expected. The new phases at $U<0$
naturally emerge through the 
commensurate-incommensurate transition.\cite{boz}

We shall also show that, regardless of the starting point, some
properties of the system are not accessible within the effective low
energy theory. The quantity in question is the diamagnetic (or
persistent) current, which turns out not to be an infra-red
property. All electrons participate in this current. In particular,
the curvature of the single-particle spectrum at the Fermi points
becomes important, so that linearization of the spectrum, being the
usual prerequisite in the derivation of any effective low-energy
theory, completely destroys this effect. Consequently, within the
bosonization approach in the context of the fermion ladder, it is
impossible to describe the analog of the Meissner effect that can be
seen in bosonic ladders.\cite{og01} Details are presented in section
\ref{chain}.

\section{Low energy effective theory: Band basis}
\label{band}

In this Section we derive the effective low energy theory for the 
model~(\ref{ham}) taking the exact single particle spectrum as 
our starting point. As mentioned above, there exists an alternative
approach, which starts with disconnected (but interacting) chains.
The relation between the two will be discussed in the next Section.

{
\begin{figure}[ht]
\epsfxsize=9.6 cm
\centerline{\epsfbox{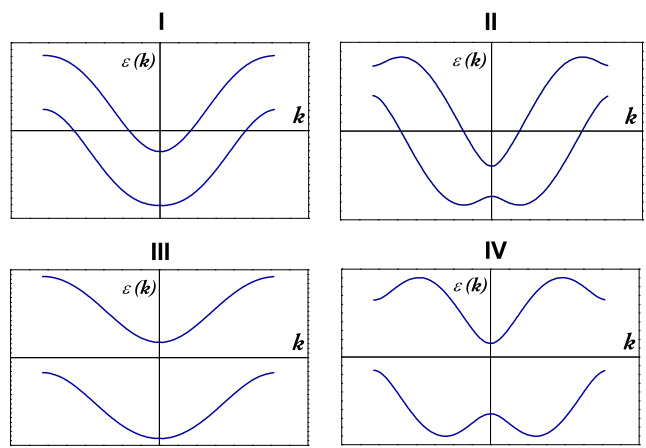}}
\caption{Possible types of the single-particle spectrum as a function
of in-chain momentum. For any given $\tau$, increasing the flux will
eventually open a band-gap in the non-interacting spectrum.  Close to
this transition one of the two bands is almost empty while the other
is almost full. At this point, curvature effects of the spectrum
become important; these are beyond the scope of this paper.}
\label{spectr1}
\end{figure}
}

\subsection{Single-particle spectrum}
\label{sps}

The single-particle part of the Hamiltonian (\ref{ham}) can be
diagonalized by the unitary transformation
\bea
c_1 (k) &=& u_k \alpha_k + v_k \beta_k, \nn
c_2 (k) &=& v_k \alpha_k - u_k \beta_k 
\label{bog}
\eea
\noindent
where the ``coherence factors'' $u_k,v_k$ [which are positive as the
signs are written explicitly in Eq.~(\ref{bog})] are given by
(if not stated otherwise, in this section we will measure the momentum $k$
in units of the inverse longitudinal lattice spacing, $1/a$)
\bea
u^2_k &=& \frac{1}{2} 
\left[ 1 - \frac{\sin k \sin \pi f}{\sqrt{\sin^2 k \sin^2 \pi f + \tau^2}}
\right] \nn
v^2_k &=& \frac{1}{2} 
\left[ 1 + \frac{\sin k \sin \pi f}{\sqrt{\sin^2 k \sin^2 \pi f + \tau^2}} 
\right].
\label{uv}
\eea
\noindent
The resulting single-particle Hamiltonian describes two
bands
\[
H_{0} = \sum_k \left[ \epsilon_\alpha \alpha_k^\dagger \alpha_k 
+ \epsilon_\beta \beta_k^\dagger \beta_k \right],
\]
\noindent
with the spectrum
\be
\epsilon_{\alpha (\beta)} = -2 t_0 \left[
\cos k \cos \pi f \pm \sqrt{\sin^2 k \sin^2\pi f + \tau^2} \right].
\label{sp}
\ee
\noindent
In the absense of the flux the coherence factors are independent of
momentum ($u^2=v^2=1/2$) and the Hamiltonian $H_0$ consists of the
usual symmetric and antisymmetric bands each with the
cosine spectrum, split by $2t_\perp$.

In the presence of the flux the spectrum can take one of the four
typical shapes depending on the value of $\tau$ and the flux. These
are illustrated in Fig.~\ref{spectr1}. For completeness we include a
detailed discussion of the properties of the spectrum as a function of
flux and interchain hopping in Appendix~\ref{spectrum_appendix}.

The half-filled ladder is characterized by zero chemical
potential. When the spectrum (\ref{sp}) takes the form depicted in the
two bottom graphs in Fig.~\ref{spectr1}, the system at $\mu = 0$ is a
band insulator. In that case interaction effects (as well as the
external field) are not expected to drastically change the
nature of the ground state of the non-interacting system. We will not
discuss that case in the present paper.

The top two graphs in Fig.~\ref{spectr1} describe the ``metallic''
phase of the non-interacting system. In this case at half filling both
bands are partially filled and each band is characterized by its own
Fermi momentum $k_F^{\alpha(\beta)}$ satisfying $k_F^\alpha+k_F^\beta
= \pi$. In what follows we will use the notation $k_F\equiv k_F^\alpha$ 
(so that $k_F^\beta=\pi-k_F$) with
\be
\cos k_F = \sqrt{\sin^2 \pi f + \tau^2}.
\ee

In the presence of the magnetic flux there exists a finite
diamagnetic current in the ground state of the system. The current
operator along the oriented link between sites $n$ and $n+1$ of the
chain $\sigma$ is
\be
j_{n,\sigma} = -i t_0 
\left( e^{i\pi f\sigma} c_{n,\sigma}^\dagger c_{n+1,\sigma} 
- h.c. \right).
\ee
This current flows in opposite directions on the two legs of the
ladder, so that the total current $j_{tot}=j_{n,+} + j_{n,-}$ will
have zero expectation value, while the expectation value of the
relative current $j_{rel} = j_{n,+} - j_{n,-}$ (in the absence of
interaction) is given by
\bea
&&\la j_{rel} \ra=  -2t_0 \sin\pi f \int \frac{dk}{2\pi} 
\Bigg\{
\cos k \left[ n_\alpha(k) + n_\beta(k) \right]
\nn
&&\quad\quad\quad\quad\quad\quad
-\frac{\sin^2 k \cos\pi f \left[ n_\alpha(k) - n_\beta(k) \right]}
{\sqrt{\sin^2 k \sin^2\pi f + \tau^2}}
\Bigg\},
\label{jrel}
\eea
where $n_{\alpha(\beta)}(k)=\la c_{\alpha(\beta)}^\dagger (k)
c_{\alpha(\beta)}(k)\ra$ are the occupation numbers of the two
bands. The current~(\ref{jrel}) is a periodic function of the flux
with a period $\Delta f=1$. In Fig~\ref{pc} we plot the flux
dependence of $\la j_{rel}\ra$ within a single period. Notice that the
current changes its sign under transformation $f\to\pi -f$; at $f=1/2$
$\la j_{\rm rel}\ra =0$ due to the recovery of time reversal symmetry
at this point.

{
\begin{figure}[ht]
\epsfxsize=7.0 cm
\centerline{\epsfbox{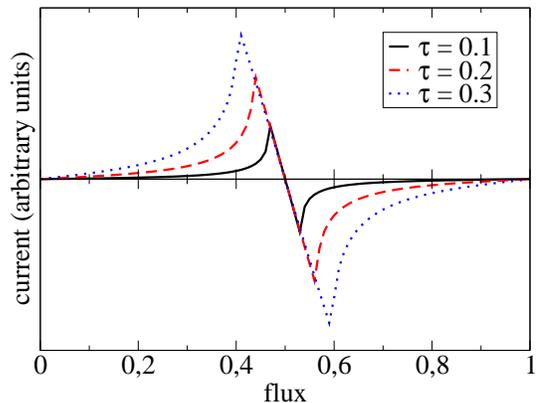}}
\caption{Diamagnetic current as a function of flux in the absence of
interaction. Only one period in $f$ is shown. The cusps correspond to
the band gap opening.}
\label{pc}
\end{figure}
}

In the limit $\tau\to 0$, when the ladder decouples into two
completely disconnected chains, the appearance of the flux in the
Hamiltonian (\ref{ham}) is a gauge artifact. Indeed, a careful
evaluation of the integral in Eq.~(\ref{jrel}) will show that $\la
j_{rel}\ra =0$ at $\tau=0$. Expanding Eq.~(\ref{jrel}) for small
$f,\tau\ll 1$ and recovering the dependence on the lattice spacing
$a$, one finds that
\be
\la j_{rel} \ra = \frac{v_F}{3a} f\tau^2 
\left[1 + O(f^2,\tau^2)\right].
\ee

Despite being small in this limit, the diamagnetic current is not an
infra-red phenomenon. Its dependence on $\tau$ is apparently the
effect of a finite curvature of the single-particle spectrum. Notice,
that as seen from Fig~\ref{pc}, $\la j_{\rm rel}\ra$ is non-zero even
in the insulating phase. Thus the current is a non-universal quantity
contributed by all electrons and not only those in the vicinity of the
Fermi points. Consequently, effects related to such a persistent current
can not be addressed in terms of any Lorentz-invariant effective
low-energy theory (we will further comment on this issue in
Section~\ref{chain}). Thus, at present we are unable to calculate the
effect of the interaction on the diamagnetic current. However, it is
clear that even in the presence of interaction the current will still
persist and all the correlation-related phenomena discussed in this
paper will coexist with it.

\subsection{Interaction Hamiltonian}
\label{int1}

Now we are going to apply the standard rules of Abelian
bosonization\cite{boz,gia} to derive the effective low-energy theory.
First, we will assume that the Fermi energy is sufficiently far from
the bottom of the $\beta$-band. Then we linearize the two-band
spectrum Eq.~(\ref{sp}) in the vicinity of the four Fermi points, $\pm
k^{\alpha}_F$ and $\pm k^{\beta}_F$. The associated low-energy degrees
of freedom are described in terms of smoothly varying chiral (right
and left) fermionic fields, $R_{\alpha(\beta)}(x)$ and
$L_{\alpha(\beta)}(x)$. This defines the continuum limit of the model
in which the non-interacting part of the lattice Hamiltonian, including
both the inter-chain hopping and the coupling to the flux, transforms
to the kinetic energy of the chiral particles:
\[
H_{0} = - i v_F \sum\limits_{\nu=\alpha,\beta}
\int dx \left( R_\nu^\dagger \partial_x R_\nu
- L_\nu^\dagger  \partial_x L_\nu  \right),
\]
\noindent
where $v_F=2t_0 a\sin k_F/\cos\pi f$ is the Fermi velocity which at
half filling is the same for both bands.

Specializing to the vicinity of the four Fermi points in the coherence
factors Eqs.~(\ref{uv}) we find the low-energy correspondence between
the original lattice operators $c_{i,\sigma}$ and the chiral fields
$R_{\nu}$ and $L_{\nu}$. Then, the interaction terms in the
model~(\ref{ham}) become
\begin{widetext}
\begin{eqnarray}
&&{H}_{int} \approx a \sum\limits_i \Bigg\{ 
g_1 \left( :J_{R\alpha}J_{L\alpha}: +:J_{R\beta}J_{L\beta}: \right)
+
g_2 \left( :J_{R\alpha}J_{R\beta}: + :J_{L\alpha}J_{L\beta}: \right)
+
g_3 \left( :J_{R\alpha}J_{L\beta}: + :J_{L\alpha}J_{R\beta}: \right) 
\nn
&&\quad\quad\quad\quad\quad
+
g_4 \left(:R_\alpha^\dagger L_\alpha R_\beta^\dagger L_\beta:+
:L_\alpha^\dagger R_\alpha L_\beta^\dagger R_\beta:\right)-
g_5\left(:R_\alpha^\dagger L_\alpha^\dagger R_\beta L_\beta:
+ : R_\alpha L_\alpha R_\beta^\dagger L_\beta^\dagger:\right)+
\nonumber\\
&& 
\nonumber\\
&& \quad\quad\quad\quad\quad
+
g_6\left(
:R_\alpha^\dagger(x_i)R_\alpha^\dagger(x_{i+1})
L_\beta(x_i)L_\beta(x_{i+1}):
+ :R_\alpha(x_i)R_\alpha(x_{i+1})
L_\beta^\dagger(x_i)L_\beta^\dagger(x_{i+1}):\right)+
\nonumber\\
&& \quad\quad\quad\quad\quad
+
g_6\left(
:L_\alpha^\dagger(x_i)L_\alpha^\dagger(x_{i+1})
R_\beta(x_i)R_\beta(x_{i+1}):
+ :L_\alpha(x_i)L_\alpha(x_{i+1})
R_\beta^\dagger(x_i)R_\beta^\dagger(x_{i+1}):\right)
\Bigg\}.
\label{hi1}
\end{eqnarray}
\noindent
where $J_{R\nu} = :R^{\dagger}_{\nu} R_{\nu}:$ and $J_{L\nu} =
:L^{\dagger}_{\nu} L_{\nu}:$ are the chiral densities of the right-
and left-moving fermions with the band index $\nu$ (the symbol ``::''
stands for normal ordering).

The first three terms in Eq.~(\ref{hi1}), characterized by coupling
constants $g_1$,$g_2$ and $g_3$, describe the density-density
interaction, whereas terms with amplitudes $g_4$, $g_5$ and $g_6$
correspond to the interchain umklapp, interchain back-scattering and
in-chain umklapp terms, respectively. Explicit expressions for the
$g_i$ in terms of the original microscopic theory is given in
Appendix~\ref{phasebd}. The coupling constants depend on the
interaction constants of the microscopic model (\ref{ham}) and,
through the coherence factors Eq.~(\ref{uv}), on the external flux.
The latter dependence plays an important role because it is
responsible for the sequence of phase transitions, described in
Section \ref{phased}), that are not accessible at $f = 0$.

The interaction Hamiltonian $H_{int}$ in Eq.~(\ref{hi1}) is the most
general form of four-fermion interaction in the band representation,
consistent with momentum conservation (modulo the reciprocal lattice
vector). All other terms contain strongly oscillating exponentials and
thus do not contribute to the low-energy theory. In particular, this
argument applies to the term
\be
g_7 e^{2i(k_F^\alpha - k_F^\beta)x}R_\alpha^\dagger L_\alpha
L^\dagger_\beta R_\beta + h.c.
\label{dk}
\ee
\noindent
Note, that if one starts building the low energy theory approximating
the ladder by two uncoupled chains, then the Fermi momenta of the two
bands are equal and the above term should be included in
Eq.~(\ref{hi1}). We will discuss this term and the relation between
the two approaches to bosonization in ladder models in
Section~\ref{chain}. However it is immediately clear, that omitting
Eq.~(\ref{dk}) from Eq.~(\ref{hi1}) can only be valid at long
distances $|x| \gg |k_F^\alpha - k_F^\beta|^{-1}$ or, equivalently,
at low energies $|\omega| \ll v_F |k_F^\alpha - k_F^\beta|$.

Now we bosonize the theory in the standard manner (our conventions are
outlined in Appendix~\ref{bosonization}). As usual, the
density-density terms (represented in Eq.~(\ref{hi1}) by $g_1$, $g_2$,
and $g_3$) renormalize the Fermi velocities and the scaling dimensions
of the vertex operators. Introducing symmetric and antisymmetric
combinations of the bosonic fields, $\phi_\pm =
(\phi_\alpha\pm\phi_\beta)/\sqrt{2}$, we diagonalize the quadratic
part of the effective bosonized Hamiltonian.  The latter is then
represented by two sine-Gordon models defined in the symmetric and
antisymmetric sectors which are coupled by the in-chain umklapp term
$g_6$:
\begin{eqnarray}
&&
{\cal H} = \frac{v_F^+}{2}\left[ K_+ \left(\p_x \theta_+\right)^2 
+ \frac{1}{K_+} 
\left( \partial_x \phi_+ \right)^2\right] -
\frac{g_4}{2\pi^2\alpha_0^2}\cos\sqrt{8\pi}\phi_+
\nonumber\\
&&
\nonumber\\
&&\quad\quad\quad\quad\quad\quad
+ \frac{v_F^-}{2}\left[ K_- \left( \partial_x \theta_- \right)^2  +  
\frac{1}{K_-} \left(\p_x \phi_-\right)^2 \right] +
\frac{g_5}{2\pi^2\alpha_0^2}\cos\sqrt{8\pi}\theta_-
-\frac{g_6}{\pi^2\alpha_0^2}\cos\sqrt{8\pi}\phi_+ 
\cos\sqrt{8\pi}\theta_-.
\label{halfbos}
\end{eqnarray}
\end{widetext}

The cosine terms in Eq.~(\ref{halfbos}), when relevant (these are the
cases $K_+ < 1$, $K_- > 1$, respectively), are responsible for a
dynamical generation of a mass gap in the corresponding sector and,
therefore, for the $U(1)$ phase transitions described in
Section~\ref{phased}.  For weak interaction, $|g_i|/\pi v_F \ll 1$,
the ``Luttinger liquid'' parameters $K_{\pm}$ are close to unity (see
Appendix~\ref{phasebd}). Consequently, the cosine terms having scaling
dimensions $2 K_+$ and $2/K_-$ in the symmetric and antisymmetric
sectors, respectively, are nearly marginal. The $g_6$ term that
couples the two sectors is therefore strongly irrelevant because of
its scaling dimension $2K_+ +2/K_- \sim 4$. The only situation when
the $g_6$ term may become important is the case when one of the
sectors is gapped, while the amplitude of the cosine term in the other
sector vanishes, i.e. either $g_4=0$ or $g_5=0$. In this case the
$g_6$ term can generate the missing cosine in the Gaussian sector and
even make the latter massive. This mechanism was recently discussed in
Ref.~\onlinecite{dtgs01} in the context of the Mott instability of a
half-filled fermionic ladder with $U=0$. Since in our model the
presence of the $g_4$ and $g_5$ terms is generic, and the lines
$g_4=0$ and $g_5=0$ characterize the phase boundaries, the only effect
of the in-chain ($g_6$) Umklapp scattering would be to modify the
equations that determine the phase boundaries without changing the
topology of the phase diagram. Being interested in the description of
distinct phases rather than their precise location, we will ignore the
$g_6$ term in the remainder of this paper.

Thus the effective low energy theory for our model Eq.~(\ref{ham})
consists of two asymptotically decoupled sectors, each being a
sine-Gordon model. In the case when a strong-coupling regime developes
in either sector, a mass gap gets generated in the spectrum, and
semiclassical solutions of the equations of motion describe locking of
the bosonic field in one of the infinitely degenerate minima of the
cosine potential. Physical quantities evaluated on such solutions may
either vanish or acquire a nonzero expectation value. The former would
mean that the quantity in question is characterized by exponentially
decaying correlations. In contrast, the latter corresponds to
long-range correlations. Since local operators of the theory have a
multiplicative structure, they can indeed serve as order parameters if
gaps are generated in both sectors simultaneously. The multiplicity of
the actual values that the order parameter would take on the
semiclassical solutions, differing by a period of the cosine
potential, determines the degeneracy of the ordered ground state. The
latter always appears to be associated with a unit cell doubling and
is $Z_2$.  The complexity of the formulae, relating the four coupling
constants in (\ref{halfbos}) to the two interaction parameters in the
original Hamiltonian (\ref{ham}), as well as the magnetic field, leads
to a rich phase diagram, as we will now demonstrate using the just
outlined strategy.

\begin{table*}
\caption{\label{op}Local operaors in the half-filled ladder}
\begin{ruledtabular}
\begin{tabular}{ccccc}
Local & Lattice & Dominant & Bosonized & Ordered\\
Operator & Definition & Component & Form & Phase\\
& & & & \\
$J_\perp$ & $-it_\perp \left[ c_1^\dagger c_2 - h.c. \right]$ & $\pi$ & 
  $\cos\sqrt{2\pi}\phi_+ \cos\sqrt{2\pi}\theta_-$ & OAF\\
$\rho_-$ & $c_1^\dagger c_1 - c_2^\dagger c_2$ & 
     $\pi$ & $\cos\sqrt{2\pi}\phi_+ \sin\sqrt{2\pi}\theta_-$ & Rel. CDW\\
$\rho_{\|,+}$ & $e^{i\pi f} c_1^\dagger(x_n)c_1(x_{n+1}) + 
  e^{-i\pi f} c_2^\dagger(x_n)c_2(x_{n+1}) + h.c.$ & $\pi$ &
  $\sin\pi f \cos\sqrt{2\pi}\phi_+ \cos\sqrt{2\pi}\theta_-$ & BDW\\
$\rho_{\|,-}$ & $e^{i\pi f} c_1^\dagger(x_n)c_1(x_{n+1}) - 
  e^{-i\pi f} c_2^\dagger(x_n)c_2(x_{n+1}) + h.c.$ & $\pi$ & 
  $\sin\sqrt{2\pi}\phi_+\sin\sqrt{2\pi}\theta_-$ & Rel. BDW \\
$\rho_+$ & $c_1^\dagger c_1 + c_2^\dagger c_2$ & $\pi$ & 
  $\tan \pi f\sin\sqrt{2\pi}\phi_+ \cos\sqrt{2\pi}\theta_-$ & CDW\\
& & $2k_F$ & 
  $\sin\sqrt{2\pi}\phi_+\cos\sqrt{2\pi}\phi_-$ & \\
$\rho_\perp$ & $c_1^\dagger c_2 + c_2^\dagger c_1$ & $2k_F$ &
  $\cos\sqrt{2\pi}\phi_+\sin\sqrt{2\pi}\phi_-$ & \\
$O_{sc}$ & $c_1 c_2$ & $\pi - 2 k_F$ &
  $ i e^{i\sqrt{2\pi}\theta_+}
    \cos [ \sqrt{2\pi}\phi_- - (\pi - 2 k_F) x]$ & \\
& & 0 & $\tan \pi f e^{i\sqrt{2\pi}\theta_+}\cos\sqrt{2\pi}\theta_-$ & \\
\end{tabular}
\end{ruledtabular}
\end{table*}

\subsection{Ordered Phases}
\label{order_parameters}

As there are four distinct Fermi points in our model, any local 
operator will contain four dominant Fourier components
\bea
\label{locop}
{\cal O}(x_n) &=& {\cal O}_0 (x)
+ (-1)^n {\cal O}_{\pi} (x)
\nonumber\\
&&
\nonumber\\
&+&  
\cos (2k_F x_n) \left[ {\cal O}_{2k_F}(x)
+ (-1)^n {\cal O}_{\pi - 2k_F}(x)\right].
\nonumber
\eea
\noindent
Here ${\cal O}_{0}$ is the smooth part of the operator ${\cal O}(x_n)$
corresponding to characteristic momentum $q \sim 0$; ${\cal O}_{\pi} $
is the staggered part contributed by momenta $q \sim \pi$, which can
originate from some inter-band pairing; the components ${\cal
O}_{2k_F}$ and ${\cal O}_{\pi - 2k_F}$ can be present due to in-band
pairing. At half filling, it is only the staggered part that can
acquire an expectation value and serve as an order parameter; however
in some of the gapless phases dominant correlations may occur at
$2k_F$ or $\pi - 2k_F$ rather than $\pi$.

Local operators of interest in the case of half-filled ladder are
listed in Table~\ref{op}, which includes the microscopic lattice
definitions and the bosonized form of the dominant Fourier components.
These are given up to multiplicative factors; we preserve, however,
prefactors proportional to the magnetic flux to make clear which
quantities do not exist in zero field limit. We also indicate the LRO
that appears when order parameters (first five operators in the
Table~\ref{op} -- which also correspond to the five ``cartoons'' in
Fig~\ref{halfphases}) acquire nonzero expectation values.

An interesting observation that can be made from Table~\ref{op} is
that the OAF and the BDW are both proportional to the same low-energy
operator, implying that the two phases coexist. However, the OAF can
exist already at zero flux whereas the BDW order parameter is
proportional to the flux (at small $f$). This coexistence can be
understood by noticing that at $f\neq 0$ the BDW order parameter is
defined in a gauge invariant way, i.e. with flux-dependent phase
factors explicitly included into its definition [we remind that we
have chosen the longitudinal Landau gauge, see Eq. (\ref{t})]. As a
result, the BDW operator describing dimerization of the two chains
with zero relative phase acquires an admixture of the staggered
relative current, proportional to the flux at $f\ll 1$. This admixture
actually represents the longitudinal part of the OAF order parameter
(which is identical to $J_{\perp}$ by current conservation).  The very
appearance of such an admixture is a consequence of the \emph{explicit}
breakdown of time reversal symmetry, caused by the external flux,
which is superimposed on the \emph{spontaneous} breakdown of this
symmetry in the OAF phase.

Another ordering in which the flux plays a crucial role is the CDW
phase. As already mentioned, the ground state of the ladder at $f=0$
and not too small $\tau$ does not display this type of LRO as the
interchain hopping tends to prevent double occupancy of the rungs. It
is a curious fact that under application of the flux this state can be
recovered due to a similar, although more subtle, admixture with a
staggered flux phase. Indeed, the bosonized low-energy projection of
the operator $\rho_+$ has the form $f J_{\rm diag}$, where the
operator
\be
J_{\rm diag} \sim i  (-1)^n \sum_\sigma \sigma\left(
c_{n,\sigma}^\dagger c_{n+1,-\sigma} - h.c. \right),
\ee
represents an order parameter for an OAF state with local currents
effectively flowing across the diagonals of the plaquettes.\cite{ner}
The scalar nature of the CDW under time reversal is not violated for
the reason already mentioned in the preceeding paragraph.

Let us now turn to the derivation of the phase diagram. For the
cosine terms to become relevant and generate a gap in the spectrum,
the ``Luttinger liquid'' parameters $K_+$ and $K_-$ should be smaller
and larger than $1$, respectively. According to the definition of
$K_{\pm}$, Eqs.(\ref{kgs}), this translates into the following
conditions on the parameters of the theory:
\begin{equation}
g_1 + g_3 > 0 ; \quad g_1 - g_3 < 0.
\label{gk}
\end{equation}
Strictly speaking, these conditions are valid only to first order in
the Kosterlitz-Thouless RG equations. When the couplings $g_i
~(i=1,3,4,5)$ are all of the same order, there are important
renormalizations of the parameters $K_{\pm}$ emerging in the
second-order\cite{boz}. This means that the exact positions of the
phase boundaries depend also on $g_4$ and $g_5$. These corrections,
however, do not cause qualitative changes in the overall phase diagram
and can therefore be neglected in the leading order. For this reason,
when drawing conclusions on relevance or irrelevance of various
perturbations, we will resort to an estimation of their Gaussian
scaling dimension.

In the effective Hamiltonian~(\ref{halfbos}) there are two cosine
terms with amplitudes $g_4$ and $g_5$. Both terms have the same period
in their respective variables that defines the values of the fields
$\phi_+$ and $\theta_-$ for any semiclassical solution. Depending on
the sign of $g_4$ the field $\phi_+$ may take one of the two possible
sets of values, $ \varphi_0 = n\sqrt{\pi/2}, $ or $ \varphi_\pi =
\sqrt{\pi/8} + n\sqrt{\pi/2} $ $(n=0,\pm1, \ldots)$. Similarly,
$\theta_-$ may take one of the above values depending on the sign of
$g_5$. Consequently there are four possible ordered phases.

(i) If both $g_4$ and $g_5$ are negative, then the semiclassical
solutions are $\phi_+=\varphi_\pi$ and $\theta_- =\varphi_0$. Of all
the operators listed in Table~\ref{op} only the staggered component of
the total charge density $\rho^{(s)}_+$ has a non-zero expectation
value on the above solution. Therefore, the conditions $g_{4(5)}<0$
define the charge density wave (CDW). This phase exists only in the
presence of the magnetic field, which mixes it up with an OAF phase,
as explained previously.

(ii) For $g_4>0$ and $g_5<0$ we find the orbital antiferromagnet
(OAF), since now $\phi_+=\theta_- =\varphi_0$ and the staggered
component of the inter-chain current $J_\perp$ gains the expectation
value. In contrast to the quarter-filled case \cite{us}, the OAF phase
exists even in the absence of the magnetic field.\cite{ner,boz} We
will clarify this issue in Sec.~\ref{strong}. At $f\neq 0$ the OAF
coexists with BDW, as we already mentioned.

(iii) When both interaction constants are positive the staggered
component of the relative charge density $\rho^{(s)}_-$ has an
expectation value (since in this case $\phi_+=\varphi_0$ and $\theta_-
=\varphi_\pi$). We call the corresponding ordered phase a relative
charge density wave (Relative CDW). The magnetic field has little
effect on this phase except for the exact location of the phase
boundary on the phase diagram, which is beyond the scope of this
paper.

(iv) Finally, if $g_4<0$ and $g_5>0$ then $\phi_+=\theta_-
=\varphi_\pi$ and the staggered component of the relative longitudinal
bond density $\rho^{(s)}_{\|,-}$ acquires a non-zero expectation value
yielding the relative bond density wave (Relative BDW). Although the
operator $\rho^{(s)}_{\|,-}$ (and therefore its expectation value)
does not vanish in the absense of the magnetic field the relative BDW
does not exist at $f=0$ (see Fig.~\ref{wcpd}) since the two conditions
$g_4<0$ and $g_5>0$ can be resolved only when $f>0$.

All above long-range ordered states break spontaneously translational
symmetry of the underlying lattice (period doubling) and thus are
doubly degenerate. Topological excitations in these phases
($Z_2$-kinks) carry unit charge $Q=1$, as opposed to fractional charge
$Q=1/2$ in the quarter-filled case. This follows from the definition
of the fermionic number carried by a single kink,
\bea
Q &=& \sum_{\nu=\alpha,\beta} \int_{-\infty}^{\infty} dx ~\left[ J_{R\nu} (x)+
J_{L\nu} (x) \right]\nonumber\\
&=& \sqrt{\frac{2}{\pi}}  \int_{-\infty}^{\infty} dx ~\partial_x \phi_+ (x)
\eea
and the fact that each kink interpolates between the vacuum values of
the field $\phi_+$ at $x \to \pm \infty$ that differ by a period of
the cosine potential, equal to $\sqrt{\pi/2}$.

\subsection{Non-ordered phases}
\label{nop}

In the previous subsection we have discussed the ordered phases
occuring under the conditions Eq.~(\ref{gk}), i.e. when both sectors in
the effective Hamiltonian~(\ref{halfbos}) acquire gaps. In all other
cases there exist gapless excitations. These are characterized by 
correlation functions that at large distances decay as a power law
\[
\langle\mathcal{O}(x) \mathcal{O}(0)\rangle
\sim 1/x^d,
\]
where $d$ is the scaling dimension of the operator $\mathcal{O}$.
Correlations with slowest decay are usually referred to as
dominant. In the phase diagram Fig.~\ref{wcpd} we categorize the
gapless phases according to their dominant correlations, indicating
the corresponding order parameter in parentheses. In this subsection
we briefly describe such phases.

If the conditions Eq.~(\ref{gk}) are reversed and $K_+>1$,
$K_-<1$, then both sectors are gapless and the system is a Luttinger
liquid. In this case the dominant correlation function is that of the
pairing operator $O_{sc}$ at wavevector $\pi-2k_F$:
\be 
\la O_{sc}^\dagger
(\tau,x) O_{sc} (0) \ra \sim 
\frac{\cos[(\pi-2k_F) x/a]}{|v_+\tau-ix|^{2/K_+} 
| v_-\tau - ix|^{K_-}} .  
\ee

There are two other cases when only one of the conditions Eq.~(\ref{gk})
is violated. Then only one of the sectors acquires a gap while the
other remains gapless:

(i) If $K_+<1$, $K_-<1$, then the ``charge'' sector is gapped, but the
``relative'' sector remains gapless. By formal analogy with the
Hubbard model we call this phase a Mott Insulator. In such state,
incommensurate density or bond-density correlations with
characteristic momentum $2k_F\neq\pi$ are dominant. Indeed, depending
on the sign of $g_4$, either $\cos\sqrt{2\pi} \phi_+$ or
$\sin\sqrt{2\pi}\phi_+$ acquire finite expectation values. Therefore,
either the transverse bond density $\rho_\perp$ or the $2k_F$ part of
the total charge density $\rho_+$ display slowest algebraic decay of
the corresponding correlation function determined by the ``relative''
sector (see Table~\ref{op} for bosonized expressions). So at $g_4 > 0$
\be 
\la \rho_\perp (\tau,x)
\rho_\perp (0) \ra \sim \frac{\cos (2k_F x/a) }{|v_-\tau -
ix|^{K_-}}.  
\label{rho-perp}
\ee 
If $g_4<0$, then Eq.~(\ref{rho-perp}) applies to the correlation 
function of $\rho_+$.

(ii) If $K_+>1$, $K_->1$, then the ``charge'' sector is gapless, but
the relative sector acquires a gap. By analogy with spin-gap systems,
we call such a phase a Luther-Emery liquid.\cite{lel} Now it is
$\theta_-$ that takes one of the two semiclassical values depending on
the sign of $g_5$. It turns out, however, that this phase can only
occur when $g_5<0$, so that $\la\cos\sqrt{2\pi}\theta_- \ra\ne 0$, and
the dominant correlation is that of the pairing operator at zero
momentum (with the power law determined by the ``charge'' sector)
\be
\la O_{sc}^\dagger (\tau,x)  O_{sc} (0) \ra \sim 
\frac{1}{|v_+ \tau-ix|^{1/K_+} }.
\ee

The phase boundaries as a function of $U,V,f,\tau$ can be calculated
by solving Eqs.(\ref{g12}) for when $g_4$ or $g_5$ is zero, or $K_+$
or $K_-$ is one.  For completeness, these are written in
Appendix~\ref{phasebd}. The complete weak-coupling phase diagram is
plotted in Figure \ref{wcpd} and was discussed in
Section~\ref{phased}.

\section{Low energy effective theory: Chain basis}
\label{chain}

In this section we briefly review the effective low energy theory that
one can derive taking two independent chains as a starting
point. Interchain hopping is then taken into account already at the
bosonization level similarly to interaction terms.  This approach is
valid as long as $t_\perp\ll v_F (f) a$, where $v_F (f)$ is the
renormalized velocity (see below). In the absence of the magnetic
field the chain-basis description of the spinless ladder has been
widely used in literature.\cite{boz,gia} Skipping inessential
details, below we will give a brief review which will help to analyze
differences between the two approaches and further clarify the role of
the magnetic flux.

In the chain-basis approach, one starts by linearizing the fermion
dispersion on each chain in the vicinity of the two Fermi points, $\pm
k_F = \pm\pi/2$, defines chiral fermion fields, and then expresses the
inter-chain hopping in terms of these fields. On the lattice, the
magnetic field was introduced in the Hamiltonian~(\ref{ham}) via the
Peierls substitution Eq.~(\ref{t}). Here it is convenient to split the
phase exponential in Eq.~(\ref{t}) into its real and imaginary part.
The real part contributes to the renormalization of the Fermi
velocity, $v_F \to v_F (f) = v_F \cos (\pi f)$, which is of minor
importance as long as $f$ is not too close to $1/2$. The imaginary
part can be written in terms of the densities of the left and right
particles, so that the single-particle perturbation to free chiral
fermions is of the form:
\be
\ham_1 (x) = -{\bf h}_R \cdot {\bf J}_R - {\bf h}_L \cdot {\bf J}_L
\label{h1}
\ee
where the chiral densities (vector currents) are defined as
\be
{\bf J}_R = \frac{1}{2} R^\dagger_\alpha 
\mathbf{\sigma}_{\alpha\beta} R_\beta, 
\quad\quad 
{\bf J}_L = \frac{1}{2} L^\dagger_\alpha 
\mathbf{\sigma}_{\alpha\beta} L_\beta.
\ee
Here $\s^a ~(a=x,y,z)$ are the Pauli matrices and $\alpha$ and $\beta$
are the chain indices. In Eq.~(\ref{h1}) the vector currents appear to
be coupled to the effective chiral ``magnetic'' fields
\be
{\bf h}_R = (h_\perp,0,-h_\|), \quad\quad {\bf h}_L = (h_\perp,0,h_\|).
\label{hchain}
\ee
with
\[
h_\| = 4 t_0 \sin(\pi f), \quad
h_\perp = 2t_\perp.
\]

Bosonizing Eq.~(\ref{h1}) directly one can find 
\be
\ham_1 = \frac{h_\perp}{\pi\alpha_0}
\cos\sqrt{2\pi}\phi_- \sin\sqrt{2\pi}\theta_- 
- h_\| \sqrt{\frac{2}{\pi}} \p_x \theta_-.
\label{h1b}
\ee
Notice that only the relative field, $\phi_-=(\phi_1-\phi_2)/\sqrt{2}$
and its dual, $\theta_-=(\theta_1-\theta_2)/\sqrt{2}$ appear in
Eq.~(\ref{h1b}) since we are discussing inter-chain processes. To
avoid confusion, we remind that here $\phi_-$ and $\theta_-$ are the
differences between the corresponding bosonic fields defined at each
chain. The remainder of the Hamiltonian in the relative sector
originates from the kinetic term and the interaction, so the total
Hamiltonian has the form
\be
\mathcal{H}_- = \mathcal{H}_G + \mathcal{H}_{SG} + \mathcal{H}_1,
\label{hbb}
\ee
where $\mathcal{H}_G$ is the Gaussian model with the interaction
parameter $K$ expressed in terms of $g_s=(4V-U)a$ in the standard way,
Eq.~(\ref{kgs}), and $\mathcal{H}_{SG}\propto U\cos\sqrt{8\pi}\phi_-$
is the sine-Gordon term.

The last term in Eq.~(\ref{hbb}) $\ham_1$ is the sum of a nonlocal
vertex operator and a derivative of the dual field $\theta_-$. The
latter appears only in the presence of the magnetic field and may be
interpreted as the bosonized version of the Lorentz-invariant current
$\tilde{j}\propto J^z_R-J^z_L$. It is important to realize, however,
that this is not the physical relative current already discussed in
Section~\ref{sps}. Moreover, $\tilde j$ appears not to be
gauge-invariant: its expectation value in the ground state of the
non-interacting system is non-zero even in the absence of inter-chain
hopping. In what follows we demonstrate that, physically, $\ham_1$
[Eq.~(\ref{h1b})] describes the splitting of the two Fermi points in
the chain basis into the four Fermi points introduced in
Section~\ref{sps}.

Treating Eq.~(\ref{hbb}) presents certain difficulties related to the
non-local nature of the perturbation (\ref{h1b}). Global rotations of
the quantization axis for the Abelian bosonization, which proved to be
efficient at $f=0$ (see Refs.~\onlinecite{nlk93,boz}), are not helpful
here because of the more complicated structure of ${\cal H}_1$. Let us
therefore perform a \emph{chiral} rotation which makes the
``magnetic'' fields ${\bf h}_R$ and ${\bf h}_L$ antiparallel and
aligned along the $z$-axis:
\be
{\bf h'}_R = (0,0,-h), \quad\quad {\bf h'}_L = (0,0,h),
\ee
\noindent
where $h^2=h_\perp^2+h_\|^2$. This simplifies the bosonic form 
Eq.~(\ref{h1b}) which now contains only the derivative term
\be
\ham_1 = -h \sqrt{\frac{2}{\pi}} \p_x \theta_-.
\label{h1c}
\ee

To achive this result we need to rotate the right field by $\omega$
and the left field by $-\omega$ about the $y$ axis where $\tan \omega
= h_\perp/h_\|$ (so that $\omega=\pi/2$ corresponds to the absence of
flux). In terms of the fermion operators this chiral $SU(2)$ rotation
can be written as
\be
R_\alpha \to \left[ e^{i\omega \sigma_y/2}\right]_{\alpha\beta} R_\beta,
\quad
L_\alpha \to \left[ e^{-i\omega \sigma_y/2}\right]_{\alpha\beta} L_\beta. 
\label{zrot}
\ee
The transformation (\ref{zrot}) is closely related to the Bogolyubov
transformation Eq.~(\ref{bog}) used to diagonalize the single-particle
Hamiltonian in the band approach. Below we analyze the relation
between the two in detail. 

While the rotation (\ref{zrot}) simplifies the single-particle terms
in the Hamiltonian, the interaction terms undergo a nontrivial
modification. Since the rotation involves only ``relative'' fields,
the ``charge'' sector is left unaffected and is a usual sine-Gordon
model, with a cosine term $\propto - U \cos \sqrt{8\pi} \Phi_+$. For
the rest of this Section we will assume this interaction to be strong
enough to generate the gap in the ``charge'' sector and focus on the
``relative'' degrees of freedom. The Hamiltonian density describing
the ``relative'' sector transforms to

\begin{widetext}
\bea
&&\ham_- = \frac{v_-}{2} 
\left[ K_- \Pi_-^2 + \frac{1}{K_-} (\p_x \phi_-)^2 \right] 
- \frac{g_\phi}{(\pi\alpha_0)^2} \cos \sqrt{8\pi}\phi_- 
-  \frac{g_\theta}{(\pi\alpha_0)^2} \cos \sqrt{8\pi}\theta_- 
-h \sqrt{\frac{2}{\pi}} \p_x \theta_-
\nonumber\\
&&
\nonumber\\ 
&&
\quad\quad\quad\quad\quad\quad\quad\quad
+ \frac{g_r}{\pi^{3/2}\alpha_0} 
\left[ (\p_x \phi^R_-)\sin \sqrt{8\pi} \phi_-^L  
+  (\p_x \phi^L_-)\sin \sqrt{8\pi} \phi_-^R \right],
\label{ham_z}
\eea

\end{widetext}
\noindent
with all the coupling constants listed in the Appendix~\ref{phasebd}.
In Eq.~(\ref{ham_z}) the relative sector represents a $Z_4$
model\cite{wieg,giaschulz} modified by the $g_r$ term. Notice that the
latter has zero conformal spin and scaling dimension
\[
d_r = 1 + \frac{1}{2} \left( K_- + \frac{1}{K_-} \right)>2
\]
for any $K_-\ne 1$. Thus this term is irrelevant in the RG sense.Thus,
as for the ladder in the absence of the magnetic flux, the effective
model appears to be two-cosine $Z_4$ model with a topological
term. The only but important difference with the $f=0$ case is a
nontrivial dependence of the coupling constants on the ratio
$\tau/\sin\pi f$ through the rotation angle $\omega$.

At $h = 0$, the $Z_4$ model in Eq.~(\ref{ham_z}) always displays a
strong-coupling regime in the infrared limit accompanied by a
dynamical generation of a mass gap. If $K<1$, then $g_\phi$ term is
relevant while the term $g_\theta$ with the dual field is irrelevant.
Hence the field $\phi_-$ gets locked and the term $h\p_x\theta_-$ has
no effect on the corresponding long range order. In particular if
$g_\phi >0$, then the ground state is the relative CDW, which becomes
the OAF when $g_\phi$ changes sign (e.g. due to the variation of the
flux). However, locking of the field $\phi_-$ does not prevent the
gradient of the dual field $\p_x \theta_-$ to acquire a finite
expectation value and, thus, produce a finite splitting of the Fermi
momenta proportional to $h$ (see the discussion below).

In the opposite case $K>1$, the $g_{\theta}$ term is relevant and
leads to locking of the dual field $\theta_-$.  As long as $h$ remains
smaller that the gap generated in the ``relative'' sector, the ground
state is the commensurate CDW, which replaces the MI shown in the
phase diagram \ref{wcpd}. In this phase the vacuum value of $\theta_-$
resides in one of the minima of the cosine potential and remains
spatially uniform, implying that $\la \p_x \theta_- \ra = 0$. Thus no
band splitting occurs in this regime reflecting the fact that the
chains remain effectively decoupled. This is a manifestation of the
Anderson confinement \cite{and} - the in-chain correlations suppress
single-particle tunneling between the chains in the low-energy
limit. Consequently the flux has little effect on this phase.

The situation changes when $h$ reaches a critical value $h_c$
proportional to the mass gap
\be
h_c \sim \Delta_s/ v_F a \sim  \exp (- 2\pi v_F /|g_s| ) \ll 1 
\quad  (g_s<0). 
\label{transition}
\ee
At this point a commensurate-incommensurate transition takes place,
the long-range order disapears and a finite gradient $\la \p_x\theta_s
\ra$ emerges in the ground state, following the universal square-root
increase $\sim \sqrt{h - h_c}$ slightly above the
threshold.\cite{boz} The commensurate CDW gets replaced by the MI
phase with incommensurate leading correlations. The appearance of a
finite average $\la \p_x\theta_s \ra$ indicates that the two doubly
degenerate Fermi points, characterizing the bare-particle spectrum in
the chain basis, are getting split.

Note, that the commensurate-incommensurate transition can be seen
within the band basis approach if one takes into account an additional
interaction process mentioned previously in Eq.~(\ref{dk}), which is
usually disregarded since formally it does not conserve momentum. The
term is the interband backscattering
\be
R_\alpha^\dagger R_\beta L_\beta^\dagger L_\alpha + h.c.  
\ee 
In bosonized form this can be written as
\be 
\frac{g_7}{(2\pi\alpha_0)^2}
\cos(\sqrt{8\pi} \phi_- + 2hx) 
\ee 
where the relation between $g_7$ and the microscopic parameters of the
model is given in Eq.~(\ref{g7}). Adding this term to the bosonic
two-band Hamiltonian (\ref{halfbos}) and making a shift
\be 
\phi_- \rightarrow \phi_- -
h x / \sqrt{2\pi} 
\ee 
one transforms the relative part of the Hamiltonian (\ref{halfbos}) to
the form identical to the $Z_4$ part in Eq.~(\ref{ham_z}) (up to
duality transformation $\phi_- \leftrightarrow \theta_-$). The above
analysis holds identically for this representation. While the case
$h\ll 1$ may be easier to describe within the chain approach, the band
picture should always give the correct result. In particular, to find
the above CDW order, one has to notice that for sufficiently small $h$
when $\la \cos \sqrt{2\pi}\phi_- \ra \neq 0$, the $2k_F$ component of
the CDW order parameter listed in Table~\ref{op} acquires a nonzero
expectation value (now $2k_F=\pi$), and the MI phase in the bottom
left of the phase diagram~\ref{wcpd} becomes a CDW phase with long
range order.

\section{Strong Coupling}
\label{strong}

In this Section we discuss the behavior of the system in the
strong-coupling limit, $|U|,|V| \gg t_0$. In the atomic limit when
hopping is completely neglected ($t_0 = t_{\perp} = 0$), the particles
are localized on sites, and there are four possible ground
states. Which state has the lowest energy is determined by the signs
of the interaction parameters $U$ and $V$: (i) if $U,V>0$, then the
ground state is the Relative CDW (the SDW$^z$ in the ``spin''
language) depicted in Fig.~\ref{halfphases}; (ii) when $U<0$ and $V>0$
the state is the CDW also shown in Fig.~\ref{halfphases}; (iii) in the
opposite case $U>0$ and $V<0$ all particles fully occupy one chain
keeping the other empty; and finally, (iv) when $U,V <0$ we have
complete phase separation.

Of the above four ground states the first two are accessible in the
weak coupling approach as can be seen from the phase diagram in
Fig.~\ref{wcpd}. The phases (iii) and (iv) do not have the lowest
energy when the bandwidth $t_0$ is greater or of the same order as $U$
and $V$ and thus have no analog in weak coupling.

To make further links with the weak coupling approach we now need to
take into account the hopping terms and the magnetic field. Of the above four
cases only the first one needs to be discussed in detail. Indeed, the
last two do not appear in weak coupling, while in the case (ii) we
have either doubly occupied or empty rungs, so that inter-chain
hopping and the flux do not afect the properties of the ground
state. This ground state (CDW) was discussed in the previous section
(this is the case where $h$ is smaller than the gap, see
Eq.~(\ref{transition}).

Consider the limit where $U>0$ is the largest scale in the
problem. Then we can project out states with doubly occupied
rungs. Then at half-filling and at energies well below the local
charge gap, there only remain configurations with exactly one fermion
per rung. Accordingly, the relative degrees of freedom can be
conveniently described in terms of local spin-1/2 variables using the
correspondence between two single-fermion states at a given rung $n$
and the eigenstates $| \uparrow\ra$ and $| \downarrow\ra$ of the
operator $S^z _n$.  The standard Schrieffer-Wolff transformation
\cite{sw66} leads to the following effective spin-chain model:
\bea {\cal H}_{eff} &=& J_0\sum_n \Big[
\frac{1}{2} \left( e^{2i\pi f} S_n^+ S_{n+1}^- + e^{-2i\pi f} S_n^-
S_{n+1}^+ \right) \nn && \quad\quad + \Delta S_n^z S_{n+1}^z + h_\perp
S_n^x \Big],
\label{strong1}
\eea
where $J_0=2t_0^2/U$ is the exchange constant, $\Delta = (J_0+V)/J_0$
is the anisotropy parameter, $h_\perp=t_\perp/J_0$ is the transverse
field and $f$ is the external flux in the original ladder model.
Notice, that the Schrieffer-Wolff transormation leaves the ``charge''
sector gapped, so that all subsequent analysis pertains to the
``relative'' sector.

In the spin language, the Relative CDW order parameter (see
Table~\ref{op}) corresponds to the staggered magnetization in the
$z$-direction $(-1)^n S_n^z $, while the OAF order parameter is the
staggered magnetization in the $y$ direction, $(-1)^n S_n^y $.
Magnetization in the $x$ direction corresponds to the transverse bond
density. Since the uniform transverse field $h_{\perp}$ breaks the
$U(1)$ symmetry of the XXZ chain, the staggered component
$(-1)^nS_n^x$ never acquires a non-zero expectation value.

The effective spin model Eq.~(\ref{strong1}) is not integrable and its
general solution remains unknown. Nevertheless, there exist at least
three cases where further progress can be made: (a) the case $f=0$
which has been studied previously (see e.g. Ref.~\onlinecite{cel03});
(b) the vicinity of the $SU(2)$-symmetric point $\Delta=1$ and $f\ll
1$; and (c) the vicinity of $f=1/2$. In what follows we discuss
these three cases.

(a) In the absence of the flux $f=0$ the Hamiltonian (\ref{strong1})
becomes equivalent to the transverse field XXZ model. Despite not
being integrable, much is known about such models.\cite{cel03} Here,
we summarize the results for the sake of completeness referring the
interested reader to the literature for more details.

First consider the case $V>0$ (so that $\Delta>1$) and the limit $V
\gg 2t_0^2/U$ in which the exchange anisotropy becomes very large,
$\Delta\gg 1$. In this case one can retain only the last two terms in
(\ref{strong1}), so that the Hamiltonian becomes equivalent to the
one-dimensional Ising model in a transverse magnetic field. At
$h_\perp<\Delta/2$ the ground state is the ordered Neel phase with
$\la (-1)^n S_n^z \ra\ne 0$. This ordering translates to the Relative
CDW for the original ladder, in agreement with the weak-coupling
picture. However, when $h_\perp = \Delta/2$ a $Z_2$ (Ising) transition
to a disordered phase takes place. This transition is not present in
the weak-coupling phase diagram, so we will not discuss it any
further.

Now, if $V<0$ but is sufficiently small ($-1<\Delta<1$), then in the
absence of $h_\perp$ the Hamiltonian~(\ref{strong1}) corresponds to
the critical XXZ model, which has a gapless excitation spectrum and
displays dominant antiferromagnetic fluctuations in the $xy$-plane.  A
non-zero transverse field breaks the $U(1)$ symmetry down to $Z_2$
partially polarizing the spins in the $x$-direction, so that $\la
(-1)^n S_n^y \ra$ develops an expectation value. This is the OAF phase
already discussed in connection with the weak-coupling phase diagram.

Finally, if $\Delta<-1$ then the model is the easy axis XXZ
ferromagnet with $\la S_n^z \ra\ne 0$. Such a ground state corresponds
to the case (iii) above, i.e. all particles localized on one of the
chains.

(b) We now consider the case $|\Delta - 1|\ll 1$ and $f\ll 1$. In
this limit the Hamiltonian (\ref{strong1}) can be represented (to
lowest order in $f$) in the form
\bea
\ham &=& J_0 \sum_n  \big\{ \vec{S}_n \cdot \vec{S}_{n+1}
+ (\Delta-1)  S_n^z S_{n+1}^z \nn
&+& h_\|  (\vec{S}_n \times \vec{S}_{n+1})^z + h_\perp S_n^x \big\},
\label{strong4}
\eea 
i.e. we have a model of a weakly anisotropic XXZ spin chain in a
magnetic field $h_\perp=t_\perp/J_0$ along the $x$ direction, and also
perturbed by the term proportional to the $z$-component of the
spin-current $({\bf S}_n \times {\bf S}_{n+1})^z$ with an amplitude
proportional to the flux, $h_\|=\pi f$. Bosonizing around the
$SU(2)$-symmetric point in the standard way,\cite{boz} one can show
that the last two terms in Eq.~(\ref{strong4}) can be written in the
form Eq.~(\ref{h1}) with the chiral ``magnetic'' fields,
Eq.~(\ref{hchain}), renormalized by the new ``bandwidth'' $J_0$. Thus
we find that the bosonized form of Eq.~(\ref{strong4}) has the
structure of Eq.~(\ref{hbb}), where the first two terms
$\mathcal{H}_G$ and $\mathcal{H}_{SG}$ constitute the Abelian bosonic
representation of the $SU(2)$-symmetric Heisenberg model in the
scaling limit, with the Lutting parameter and the effective coupling
constant renormalized by the weak anisotropy term $|\Delta -1|\propto
V$ in Eq.~(\ref{strong4}).

Notice that the resulting Hamiltonian is basically the same as that
obtained at weak coupling in the chain basis (\ref{hbb}). Therefore
the next step, namely the rotation Eq.~(\ref{zrot}) and its result
Eq.~\ref{ham_z}, can be performed in exactly the same manner as
before. As a result, at least in the region $U>0$, all conclusions
drawn at weak coupling also hold true in the strong coupling limit. In
particular the region of the phase diagram discussed in the previous
Section survivies (up to renormalizations of the phase boundaries) in
the strong coupling regime as well.

(c) Finally, we consider the case where flux is close to one half.
Here we are close to the transition to a band insulator. This region
of the phase diagram can not be well treated in our weak coupling
approach. In this sense, the strong coupling arguments compliment the
weak coupling picture presented in Section~\ref{band}.

Consider a gauge transformation
\be
S_n^+ \rightarrow S_n^+ e^{2i\pi f n}, 
\quad\quad S_n^- \rightarrow S_n^+ e^{-2i\pi f n},
\ee
which transforms the Hamiltonian (\ref{strong1}) to 
\bea
{\cal H}_{eff} &=& J_0\sum_n 
\Big[  \frac{1}{2} 
\left(  S_n^+ S_{n+1}^- + S_n^- S_{n+1}^+ \right)
+ \Delta S_n^z S_{n+1}^z
\nn
&& 
+ h_\perp \left( S_n^+ e^{2i\pi f n} + S_n^- e^{-2i\pi f n} \right)   
\Big].
\label{strong2}
\eea
The model (\ref{strong2}) is completely equivalent to
Eq.~(\ref{strong1}) but now the transverse field is non-uniform. The
situation significantly simplifies when $|f-1/2| \ll 1$. In the case
this field becomes almost staggered and can be directly bosonized (in
the region $|\Delta|<1$, i.e $V<0$).\cite{boz} As a result,
Eq.~(\ref{strong2}) becomes
\bea
\label{strong3}
H &=& \frac{u}{2}\int dx 
\left[ K \Pi^2 + \frac{1}{K} (\partial_x \Phi)^2 \right] 
\\
&+& 
h_\perp\int dx \cos \left[ \sqrt{\pi} \Theta(x) - 2\pi \left(
\frac{1}{2}-f\right) 
\frac{x}{a} \right],
\nonumber
\eea
where the Luttinger liquid parameter is given by
\be
K = \pi/2(\pi-\cos^{-1} \Delta).
\label{Kba}
\ee

For $f=1/2$, the cosine term is relevant and generates a gap in the
spectrum. The resulting ground state is the band insulator and the gap
corresponds to that seen already in the single-particle problem in
Section~\ref{sps}. This can be seen from the fact that $\cos(\sqrt \pi
\Theta)$ (which gains a non-zero expectation value at $f=1/2$)
corresponds to the {\em uniform} bond density. 

As $f$ decreases away from $1/2$, eventually the gap will close via a
commensurate-incommensurate transition. The system will now have
gapless excitations in the relative sector (i.e as in a MI), in
qualitative agreement with the weak coupling phase diagram
Fig~\ref{wcpd}.

Finally, if $V>0$ so that $\Delta>1$ one can still bosonize the
Hamiltonian~(\ref{strong2}), although now there is an extra term
proportional to $\cos{\sqrt{16\pi}\Phi}$ which is relevant (formally,
this is the case $K<1/2$ which is not captured Eq.~(\ref{Kba})). We
now have two competing cosine terms in the Hamiltonian,
$\cos\sqrt{16\pi}\Phi$ and $\cos[ \sqrt{\pi} \Theta(x) - 2\pi(1/2-f)
x/a ]$. The dual field term has the smallest scaling dimension and
therefore, when $f=1/2$, determines the character of the ground
state. If the flux is decreased, then again the order is destroyed. In
this case, however, the other relevant operator in the problem,
$\cos\sqrt{16\pi}\Phi$ should at this point acquire a non-zero
expectation value. This is the transition between the band-insulator
and the relative CDW, again in agreement with Fig~\ref{wcpd}.

\section{Summary}

We have investigated a model of interacting spinless fermions hopping
on a two-leg ladder in the presence of an external magnetic field at
half-filling. Using bosonization techniques, we constructed the
effective low energy theory where the coupling constants acquired
non-trivial dependence on the external flux. Consequently the flux
results in several phase transitions shown in the weak-coupling phase
diagram Fig.~\ref{wcpd}, i.e. BKT transitions between ordered and
disordered phases and $U(1)$ transitions between different ordered
phases.

Furthermore, we extended our weak-coupling picture by the special
consideration of the case of weakly coupled chains at small flux. We
solved the corresponding effective theory using the chiral rotation
Eq.~(\ref{zrot}). As a result we described the
commensurate-incommensurate transition from the MI phase shown in
Fig.~\ref{wcpd} at $U<0$ to the CDW. This transition happens when the
parameter $\sqrt{\tau^2+\sin^2\pi f}$ becomes small enough [see
Eq.~(\ref{transition})].
 
The weak coupling analysis is complimented by the strong coupling
arguments. In particular we showed that in some cases (in particular,
for small flux and small in-chain interaction $V$), that the
weak-coupling approach and strong-coupling approach lead to identical
low-energy theories.

Finally, we discussed the persistent (diamagnetic) current flowing in
the ladder when the external magnetic field is applied. We showed that
even in the limit of small $f$ and small $t_\perp$, all electrons in
the system contribute to this current and therefore it is a
non-universal feature which can not be described in the traditional
field-theoretic approach. We speculate that it may be an interesting
physical quantity to investigate within the non-linear bosonization
scheme.\cite{nlb}

\acknowledgments

We acknowledge useful discussions with A.M.Tsvelik, A.A.Gogolin, and
M.Fabrizio.

\appendix

\section{Relation between parameters and phase boundaries}
\label{phasebd}

Here we list the effective coupling constants for the effective
low energy theory and their relation to the microscopic 
parameters in the original ladder Hamiltonian Eq.~(\ref{ham}).

First, we show the parameters of the $g-$ology for the effective
interaction Eqs.~(\ref{hi1}) and (\ref{dk}) in Section~\ref{int1}:

\begin{subequations}
\begin{equation}
g_1= \frac{\cos^2\pi f - \tau^2}{\sin^2\pi f +\tau^2}
\left[Ua 
\tan^2 \pi f +
2 Va \frac{\tau^2}{ \cos^2\pi f}\right];
\end{equation}
\begin{equation}
g_2 =Ua +
2Va \frac{\tau^2}{ \cos^2\pi f};
\end{equation}
\begin{eqnarray}
&&g_3= 2Va + \frac{1}{\sin^2\pi f +\tau^2}
\Bigg[
\frac{Ua\tau^2}{\cos^2\pi f}
\\
&&
\quad\quad\quad\quad\quad\quad
+ 
2Va
(\sin^2 \pi f - \tau^2\tan^2\pi f)
\Bigg];
\nonumber
\end{eqnarray}
\begin{equation}
g_4 = g_2; 
\end{equation}
\begin{equation}
g_5 = g_1;
\end{equation}
\begin{equation}
g_6 = 
\frac{Va}{4}
\left(1+\tan^2 \pi f\frac{\cos^2\pi f - \tau^2}{\sin^2\pi f +\tau^2}\right)
\end{equation}
\be 
g_7 = - \frac{\tau^2}{\cos^2\pi f} \Bigg[
2Va +
\frac{Ua-2Va}{\sin^2\pi f + \tau^2}
\Bigg]
\label{g7}
\ee 
\label{g12}
\end{subequations}

\noindent
The two ``Luttinger parameters'' are then given by
\begin{equation}
K_\pm = \left[ 
\frac{1-(g_1\pm g_3)/4\pi v_F^\pm}{1+(g_1 \pm g_3)/4\pi v_F^\pm}
\right]^{1/2}
\end{equation}

The phase boundaries in Fig.~\ref{wcpd} (within the accuracy of
one-loop renormalization group approach) are given by the following 
conditions:

\bea
g_1+g_3=0 &:& \frac{U}{2V} = - \frac{2\cos^2 \pi f - \tau^2}{\cos^2 \pi f} \nn
g_1-g_3=0 &:& \frac{U}{2V} = 
\frac{\sin^2\pi f(2\cos^2\pi f - \tau^2) + \tau^4}
{\sin^2 \pi f ( \cos^2 \pi f - \tau^2) - \tau^2} \nn
g_4=0 &:& \frac{U}{2V} = -\frac{\tau^2}{\cos^2 \pi f} \nn
g_5=0 &:& \frac{U}{2V} = -\frac{\tau^2}{\sin^2 \pi f}
\eea
Notice that the boundaries depend only on the ratio $U/V$.

Finally, we list the effective constants for the effective low
energy theory in the ``chain basis'' Eq.~(\ref{ham_z})

\bea
&&g_\phi = \frac{1}{2}
\left[ 2Va \sin^2 \omega + Ua \cos^2 \omega \right] ;
\nn
&&g_\theta = \frac{2Va-Ua}{2}
\sin^2 \omega   ;
\nn
&&
g_r = (2Va - Ua) \sin2\omega; 
\nn     
&&
g_s = 
\left[4Va \cos^2 \omega + Ua (\sin^2 \omega - \cos^2 \omega)\right], \nn
&&
K_s = \left[ 
\frac{1-g_s/2\pi v_s}{1+g_s/2\pi v_s}
\right]^{1/2}.
\label{kgs}
\eea


\section{Single Particle Properties}
\label{spectrum_appendix}

Here we point out trivial properties of the single-particle spectrum
(\ref{sp}) for the sake of completeness. The upper band has its
absolute minimum at $k=0$ with
\begin{equation}
(\epsilon_\beta)_{min} = \epsilon_\beta(0) = \tau-\cos\pi f.
\label{bmi}
\end{equation}
\noindent
The lower band has its absolute maximum at $k=\pm\pi$ 
\begin{equation}
(\epsilon_\alpha)_{max} = \epsilon_\alpha(\pi) = \cos\pi f - \tau.
\label{ama}
\end{equation}
\noindent
Thus, if $\cos\pi f < \tau$, the spectrum exhibits a single-particle
gap and the non-interacting system is a band insulator at half filling.

{
\begin{figure}[ht]
\epsfxsize=7 cm
\centerline{\epsfbox{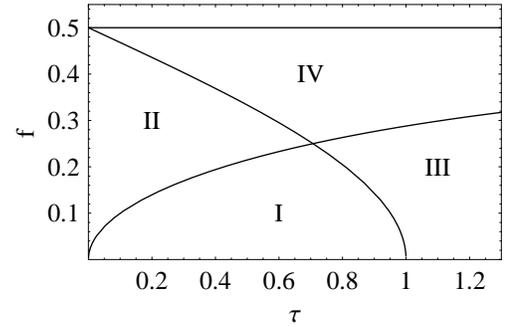}}
\caption{The phase diagram of the non-interacting system. The vertical
axis is the external flux and the horizontal axis is the ration
between inter-chain and in-chain hopping amplitudes.
Fig.~\ref{spectr1} demonnstrates the four typical shapes of the
spectrum corresponding to the four corners in which the phase diagram
is separated by the two lines Eq.~(\ref{phd}).}
\label{spectr2}
\end{figure}
}

If the flux is not too small then there is a non-trivial extremal point
$\cos k_1 = - \cot\pi f \sqrt{\tau^2 + \sin^2\pi f }.$
For the upper band the energy $\epsilon_\beta (k_1)$ has to be
compared to the remaining extremum $\epsilon_\beta (\pi)$. If
$\sin^2\pi f > \tau\cos\pi f$ the band acquires a double-well shape as
shown in plots II and IV in Fig.~\ref{spectr1} and $k_1$ turns out to
be the absolute maximum for the uper branch of the spectrum, while
$k=\pm\pi$ are local minima. For the lower branch of the spectrum and
under the same condition, $k_1$ is the absolute minimum, while $k=0$ is
the local maximum with the energy value
\begin{equation}
\epsilon_\alpha(0) = -\tau-\cos\pi f
\label{eaz}
\end{equation}
The two aforementioned conditions, namely,
\begin{equation}
\cos\pi f_2 = \tau ; \quad\quad \sin^2\pi f_1 =
\tau\cos\pi f_1
\label{phd}
\end{equation}
define the two boundaries in the phase diagram of the non-interacting
system shown in Fig.~\ref{spectr2}. These two lines separate the phase
diagram of the non-interacting system (Fig.~\ref{spectr2}) into four
parts, where the spectrum has one of the four shapes shown in
Fig.~\ref{spectr1}.


\section{Bosonization conventions}
\label{bosonization}

Here for completeness we define our bosonization conventions.  
The chiral bosonic fields are introduced via the correspondence
\begin{eqnarray}
&& \left(
R_{\nu} (x),
L_{\nu} (x)
\right)
\to
\left(\kappa_\nu/\sqrt{2\pi\alpha_0}\right)
e^{{\pm i\sqrt{4\pi} \phi^{R(L)}_\nu}},
\label{RL-bos}
\\
&& 
J_{R\nu} = \p_x \phi_{R\nu}/\sqrt{\pi}, \quad
J_{L\nu} = \p_x \phi_{L\nu}/\sqrt{\pi}.
\label{curr-bos}
\end{eqnarray}
\noindent
Here $\nu = \alpha,\beta$,  $\alpha_0$ is an ultraviolet cutoff in the bosonic theory,
$\kappa_{\nu}$ are Klein factors that ensure proper anticommutation
relations between the fermionic fields with different band indices
in representation (\ref{RL-bos}).
The $\kappa_{\nu}$ satisfy
\begin{equation}
\{ \kappa_\mu, \kappa_\nu \} = 2\delta_{\mu\nu}; \quad
\kappa_\alpha \kappa_\beta = -i.
\end{equation}
\noindent
In addition, we impose a nontrivial commutation relation between the right
and left bosonic fields belonging to the same band:
\begin{equation}
\left[ \phi^R_\mu, \phi^L_\nu \right] =  i \delta_{\mu\nu}/4.
\end{equation} 
The left- and right-moving fields can be combined into the field
$\phi$ and its dual counterpart $\theta$
\begin{equation}
\phi_\alpha = \phi_\alpha^R + \phi_\alpha^L ; \quad
\theta_\alpha = \phi_\alpha^R - \phi_\alpha^L,
\end{equation}
\noindent
with $\Pi = -\p_x \theta_{\nu}$ being the momentum conjugate to $\phi_{\nu}$.
The linear combinations
\[
\phi_\pm = (\phi_\alpha \pm \phi_\beta)/\sqrt{2}; \quad
\theta_\pm = (\theta_\alpha \pm \theta_\beta)/\sqrt{2}.
\]
describe collective bosonic degrees of freedom and the symmetric and
antisymmetric sectors of the theory.


\end{document}